\documentclass[aps,pra,twocolumn,amsmath,amssymb,nofootinbib,showpacs,superscriptaddress]{revtex4-2}
\usepackage[english]{babel}
\usepackage{latexsym}
\usepackage{graphicx}
\usepackage{epstopdf}
\usepackage{epsfig}
\usepackage{color}
\usepackage{bm}
\usepackage{amsmath}
\usepackage{amssymb}
\usepackage{amsthm}
\usepackage{dcolumn}
\usepackage{bm}
\usepackage{float}
\usepackage{hyperref}
\usepackage{color}
\usepackage{cleveref}
\usepackage[svgnames,dvipsnames]{xcolor}
\usepackage{enumerate}
\usepackage{braket}

\hypersetup{hidelinks,colorlinks=true,allcolors=DarkBlue}

\DeclareMathOperator{\Tr}{Tr}

\newtheorem*{theorem}{Theorem}

\theoremstyle{remark}
\newtheorem{remark}{Remark}

\renewcommand*{\Re}{\mathop{\mathrm{Re}}\nolimits}
\renewcommand*{\Im}{\mathop{\mathrm{Im}}\nolimits}

\begin{document}

%\preprint{APS/123-QED}

\title{Unified Gorini--Kossakowski--Lindblad--Sudarshan quantum master equation\\ beyond the secular approximation
}

\author{Anton Trushechkin}
\affiliation{Steklov Mathematical Institute of Russian Academy of Sciences, Moscow 119991, Russia}
\affiliation{National University of Science and Technology MISIS, Moscow 119049, Russia}

\email{trushechkin@mi-ras.ru}

\date{\today}

\begin{abstract}

Derivation of a quantum master equation for a system weakly coupled to a bath which takes into account nonsecular effects, but nevertheless has the mathematically correct Gorini--Kossakowski--Lindblad--Sudarshan form (in particular, it preserves positivity of the density operator) and also satisfies the standard thermodynamic properties is a known long-standing problem in theory of open quantum systems. The nonsecular terms are important when some energy levels of the system or their differences (Bohr frequencies) are nearly degenerate. We provide a fully rigorous derivation of such equation based on a formalization of the weak-coupling limit for the general case. 

\end{abstract}

\maketitle

\section{Introduction} 

Quantum master equations are at the heart of theory of open quantum systems \cite{AL,BP,RH}. They describe the dynamics of the reduced density operator of a system interacting with the environment (``bath'') and are widely used in quantum optics, condensed matter physics, charge and energy transfer in molecular systems, bio-chemical processes \cite{MK}, quantum thermodynamics \cite{KosloffQTD,AlickiKosloffQTD}, etc. The Redfield and Davies quantum master equations are well-known microscopically derived equations for a system weakly coupled to a bath and are crucial for understanding many physical phenomena. 

The Davies quantum master equation \cite{Davies,Davies2} is derived in a mathematically rigorous way and has the Gorini--Kossakowski--Lindblad--Sudarshan (GKLS) form \cite{GKS,L,BriefGKLS}, which guarantees that the corresponding dynamics of the reduced density operator is well defined. The Davies equation also satisfies a number of properties important for thermodynamics: stationarity of the Gibbs state, the detailed balance condition, a covariance law (related to the first law of thermodynamics \cite{Lostaglio,MarvianSpek, DannKosloff}), non-negativity of the entropy production (i.e., the second law of thermodynamics) \cite{AL,KosloffQTD,AlickiKosloffQTD,SpohnLeb,Gorini,Spohn1978}.

However, this equation assumes that all distinct Bohr frequencies (differences between the energy levels) of the system are well separated from each other. In other words, all differences between distinct Bohr frequencies are much higher than the dissipation rates (the secular approximation). This assumption is not satisfied if  some energy levels or Bohr frequencies are nearly degenerate (but not exactly degenerate), which is often the case in physical systems. 

The Redfield master equation \cite{Redfield} does not adopt the secular approximation and thus is more general. It is widely used in various physical applications, the role of the nonsecular terms is studied, e.g., in Refs.~\cite{IFlRedf,Brumer,Lovett}. But, unfortunately, it is not of the GKLS form and, in particular, does not preserve positivity of the density operator, thus leading to unphysical predictions. Also, this equation does not have the mentioned  thermodynamic properties.

Derivation of a mathematically correct quantum master equation which takes into account nonsecular effects is actively studied. Several heuristic approaches turning the Redfield equation into an equation of the GKLS form without the (full) secular approximation have been proposed. One possibility is the time coarse graining  \cite{Alicki,SchallerBrandes,BreuerLidar,CresserFacer,RivasRef,FarinaGio}.  Another method is a partial secular approximation followed by the approximation of slow variation of the spectral density \cite{PtaEsp,Bondar,NathanRudner,CSP}. Also, in certain cases, the so-called local approach is considered as an alternative to the secular approximation and is largely debated \cite{BreuerLoc,RHP,Hofer,Gonzalez,Correa,Decordi,CattaneoLocGlob,TruBashJETP,
TruBashScr}.
But, like the Redfield equation, these equations do not satisfy all the mentioned thermodynamic properties.  Moreover, it has been explicitly shown that the local master equation violates the second law of thermodynamics \cite{LevyKosloff}.

In this paper, we derive a unified master equation for the weak-coupling  regime in a mathematically rigorous and systematic way, which leads to the GKLS form and all the desired thermodynamic properties. The derivation is based on a rigorous formalization of the weak-coupling limit for the general (nonsecular) case. The unified equation has a simple and intuitive structure similar to that of the Davies equation. 

Interestingly, this equation coincides with the refined (thermodynamically consistent) form of the local master equation when the local approach is expected to be valid. Thus, we rigorously justify the correct form of the local master equation (popular due to its simplicity) and complete the results of Ref.~\cite{TrushVol}.

Note that the general idea was also proposed in Ref.~\cite{AL} and  master equations for a particular system were rigorously derived in Ref.~\cite{DaviesAtomRad}. General properties of the dynamics of particular models of open quantum systems with nearly degenerate spectrum were also rigorously established in Refs.~\cite{MerkliSong,MerkliSongBerman}. Here we derive a master equation for the general situation.

\section{Redfield equation and secular approximation} 

Consider the system-bath Hamiltonian 
\begin{equation}\label{EqH}
H=H_S+H_B+\lambda H_I,
\end{equation}
where $H_S$ is the isolated system Hamiltonian, $H_B$ is the isolated bath Hamiltonian, $H_I=\sum_\alpha A_\alpha\otimes B_\alpha$ is the interaction Hamiltonian (with $A_\alpha$ being system operators and $B_\alpha$ being bath operators), and $\lambda$ is a formal small dimensionless parameter. Let $H_S$ have a purely discrete spectrum: $H_S=\sum_j\varepsilon_jP_j$, where $\varepsilon_j$ are the distinct eigenvalues and $P_j$ are the corresponding eigenprojectors. The differences $\varepsilon_{j'}-\varepsilon_j$ are called the Bohr frequencies. Denote $\mathcal F$ the set of all (positive, negative, and zero) Bohr frequencies.

Let the initial system-bath state be $\rho_0\otimes\sigma_B$, where $\rho_0$ is the initial state of the system and $\sigma_B$ is a reference state of the bath such that $e^{-iH_Bt}\sigma_Be^{iH_Bt}=\sigma_B$ for all $t$ (e.g., a thermal state). The dynamics of  the reduced density operator of the system is given by $$\rho(t)=\Tr_B\big[e^{-iHt}(\rho_0\otimes\sigma_B)e^{iHt}\big],$$ 
where $\Tr_B$ is the partial trace over the bath. The density operator in the interaction picture is $
\tilde\rho(t)=e^{iH_St}\rho(t)e^{-iH_St}.$

The derivation of the quantum master equation is based on the idea of the separation of different time scales. In the rigorous derivation, this intuition is formalized by the Bogolyubov--van Hove limit: $\lambda\to0$, $t\to\infty$, $\lambda^2t=\tau={\rm const}$ (see Refs.~\cite{RH,Davies,Davies2,Spohn1980}). The standard ``physical'' derivation \cite{BP,RH} leads to the Redfield equation (we express it in a GKLS-like form \cite{FarinaGio}):
\begin{eqnarray}
&&\frac{d}{dt}\tilde\rho(\tau)=-i[\overline H_{\rm LS}(\tau),\tilde\rho(\tau)]
+\sum_{\omega,\omega'\in\mathcal F}\sum_{\alpha,\beta}
\gamma_{\alpha\beta}(\omega,\omega')\nonumber\\
&&\times e^{i(\omega'-\omega)\frac{\tau}{\lambda^2}}
\Big(A_{\beta\omega}\tilde\rho(\tau)A_{\alpha\omega'}^\dag-
\frac12\big\{A_{\alpha\omega'}^\dag A_{\beta\omega},\tilde\rho(\tau)\big\}
\Big),\qquad\label{EqRedfieldGKLS}
\end{eqnarray}
where
\begin{equation}\label{EqHLSRedf}
\overline H_{\rm LS}(\tau)=\sum_{\omega,\omega'\in\mathcal F}\sum_{\alpha,\beta}
S_{\alpha\beta}(\omega,\omega')
e^{i(\omega'-\omega)\frac{\tau}{\lambda^2}}
A_{\alpha\omega'}^\dag A_{\beta\omega},
\end{equation}
(the subindex LS stands for the Lamb shift),
\begin{gather}
A_{\alpha\omega}=\sum_{j,j'\colon\varepsilon_j-\varepsilon_{j'}=\omega}
P_{j'} A_\alpha P_{j},\:
[H_S,A_{\alpha\omega}]=-\omega A_{\alpha\omega},\label{EqAomega}
\\
\begin{aligned}
\gamma_{\alpha\beta}(\omega,\omega')&=
\Gamma_{\alpha\beta}(\omega)+\Gamma^*_{\beta\alpha}(\omega'),\\
S_{\alpha\beta}(\omega,\omega')&=
\frac1{2i}\left[\Gamma_{\alpha\beta}(\omega)-\Gamma^*_{\beta\alpha}(\omega')
\right]
\end{aligned}\nonumber
\\
\Gamma_{\alpha\beta}(\omega)=
\int_0^\infty ds\,e^{i\omega s} C_{\alpha\beta}(s)
\end{gather}
(we assume that the last integral converges). Here
\begin{equation}\label{EqCorr}
C_{\alpha\beta}(s)=
\Tr [e^{-iH_Bs}B_\alpha^\dag e^{iH_Bs}B_\beta\sigma_B]
\end{equation}
are bath correlation functions.

The matrix $\gamma_{\alpha\beta}(\omega,\omega')$ [with two double indices $i=(\alpha,\omega')$ and $j=(\beta,\omega)$] is, in general, not positive semidefinite, so, the equation is not of the GKLS form and defines the dynamics which violates positivity. 

As $\lambda\to0$, the exponents $e^{i(\omega'-\omega)\tau/\lambda^2}$ for $\omega\neq\omega'$ rapidly oscillate and the corresponding terms can be neglected (the secular approximation). After this, the equation becomes of the first standard GKLS form \cite{AL,BP} since, for each $\omega$, the matrix $\gamma_{\alpha\beta}(\omega,\omega)\equiv\gamma_{\alpha\beta}(\omega)$ (with the indices $\alpha$ and $\beta$) is well known to be positive semidefinite. We will refer to this equation as the secular or the  Davies master equation.

However, as pointed out above, the secular approximation is not always valid. If so, the formal mathematical limit in the present form does not correspond to physics. There is no limit in a concrete physical system, but all physical quantities have concrete values. The limit $\lambda\to0$ is just a mathematical expression of the fact that the dissipative dynamics caused by the coupling to the bath is much slower than all other time scales, but this is not the case if two different Bohr frequencies $\omega$ and $\omega'$ are close to each other. Obviously, nearly degenerate energy levels is a particular case.

\section{Unified master equation} 

\subsection{Derivation}\label{SecDeriv}

If we claim that, for a given physical system, a difference $\omega-\omega'$ is small and, hence, the term $e^{i(\omega'-\omega)\tau/\lambda^2}$ is not rapidly oscillating (with respect to the time scale of the dissipation), then, in the formal derivation, the difference $\omega'-\omega$ should be treated as infinitesimal and, moreover, of the order of $\lambda^2$. This should be explicitly formalized in the mathematical language of the limit. 

Let us express the system Hamiltonian $H_S$ as
\begin{equation}\label{EqHSdecompose}
H_S=H_S^{(0)}+\lambda^2\delta H_S,
\end{equation}
where $[H_S^{(0)},\delta H_S]=0$ and all nearly degenerate Bohr frequencies in $H_S$ are exactly degenerate in $H_S^{(0)}$. In other words, all distinct Bohr frequencies of $H_S^{(0)}$ are well separated. Proportionality of the remaining part to $\lambda^2$ mathematically expresses the fact that some oscillations $e^{i(\omega'-\omega)\tau/\lambda^2}$  occur on the same time scale as the dissipation.

Due to commutativity and since $H_S^{(0)}$ may be more degenerate than $H_S$, its spectral decomposition is $H_S^{(0)}=\sum_k \varepsilon_k^{(0)}P_k^{(0)}$, where $\varepsilon_k^{(0)}$ are the distinct eigenvalues and each eigenprojector $P_k^{(0)}$ is either one of $P_j$ or a sum of several $P_j$.

Denote $\mathcal F^{(0)}$ the set of the Bohr frequencies of $H_S^{(0)}$. Then, each Bohr frequency $\omega$ of the original system Hamiltonian $H_S$ can be expressed as $\omega=\overline\omega+\lambda^2\delta\omega$, where $\overline\omega\in\mathcal F^{(0)}$ and $\delta\omega$ is a Bohr frequency of $\delta H_S$. In other words, the set of Bohr frequencies $\mathcal F$ of $H_S$ is divided into  disjoint subsets (clusters) $\mathcal F_{\overline\omega}$ of the Bohr frequencies centered around $\overline\omega\in\mathcal F^{(0)}$. The difference between any pair of Bohr frequencies from the same cluster is proportional to $\lambda^2$. Physically, the Bohr frequencies from different clusters are well separated, while those from the same cluster are not:
\begin{equation}\label{EqOmegadif}
\omega'-\omega=\overline\omega-\overline\omega'+
\lambda^2(\delta\omega'-\delta\omega).
\end{equation}
So, the exponent $\exp[i(\omega'-\omega)\tau/\lambda^2]$ is  rapidly oscillating (as $\lambda\to0$) if and only if $\omega$ and $\omega'$ belong to different clusters ($\overline\omega\neq\overline\omega'$). 

Using this, let us drop the rapidly oscillating terms from the Redfield equation (\ref{EqRedfieldGKLS}) (i.e., apply the secular approximation with respect to $H_S^{(0)}$, which is a partial secular approximation with respect to $H_S$). Also, let us perform the limit $\lambda\to0$ in the arguments of $\gamma_{\alpha\beta}$, i.e., $\gamma_{\alpha\beta}(\omega,\omega')\to\gamma_{\alpha\beta}(\overline\omega,\overline\omega)=\gamma_{\alpha\beta}(\overline\omega)$ for $\omega,\omega'\in\mathcal F_{\overline\omega}$. Then we arrive at the following master equation:
\begin{eqnarray}
\frac{d}{d\tau}\tilde\rho(\tau)&=&
-i[H_{\rm LS}(\tau),\tilde\rho(\tau)]
\nonumber\\&+&\!
\sum_{\overline\omega\in\mathcal F^{(0)}}\!
\sum_{\omega,\omega'\in\mathcal F_{\overline\omega}}
\sum_{\alpha,\beta}
e^{i(\omega'-\omega)\tau}
\gamma_{\alpha\beta}(\overline\omega)
\nonumber\\
&\times&
\Big(A_{\beta\omega}\tilde\rho(\tau)A_{\alpha\omega'}^\dag
-\frac12\big\{A_{\alpha\omega'}^\dag A_{\beta\omega},\tilde\rho(\tau)\big\}\Big),\qquad
\label{EqUni1}
\end{eqnarray}
$H_{\rm LS}(\tau)=e^{iH_S\tau}H_{\rm LS}e^{-iH_S\tau}$ ($H_{\rm LS}$ is given below). In the Schr\"odinger picture and in the original time scale $t$, the equation takes the form
\begin{equation}\label{EqUniS}
\frac{d}{dt}{\rho}(t)=
-i[H_S+\lambda^2 H_{\rm LS},\rho(t)]
+\lambda^2
\mathcal D[\rho(t)]\equiv \mathcal L[\rho(t)],
\end{equation}
\begin{equation}\label{EqD}
\mathcal D\rho=
\sum_{\overline\omega\in\mathcal F^{(0)}}
\sum_{\alpha,\beta}
\gamma_{\alpha\beta}(\overline\omega)
\Big(A_{\beta\overline\omega}\rho A_{\alpha\overline\omega}^\dag
-\frac12\big\{A_{\alpha\overline\omega}^\dag A_{\beta\overline\omega},\rho\big\}\Big),
\end{equation}
\begin{equation}\label{EqHLS}
H_{\rm LS}=
\sum_{\overline\omega\in\mathcal F^{(0)}}
\sum_{\omega,\omega'\in\mathcal F_{\overline\omega}}
\sum_{\alpha,\beta}
S_{\alpha\beta}(\omega,\omega')
A_{\alpha\omega'}^\dag A_{\beta\omega},
\end{equation}
where 
\begin{gather}
A_{\alpha\overline\omega}=\sum_{\omega\in\mathcal F_{\overline\omega}} A_{\alpha\omega}
=\sum_{k,k'\colon \varepsilon_k^{(0)}-\varepsilon^{(0)}_{k'}=
\overline\omega}P_{k'}^{(0)}A_\alpha P_k^{(0)},\label{EqAbaromega}
\\
[H_S^{(0)},A_{\alpha\overline\omega}]=
-\overline{\omega}A_{\alpha\overline\omega}.\label{EqAeigen}
\end{gather}
A rigorous result (a theorem) is given in Appendix~\ref{SecRig}. Since the matrix $\gamma_{\alpha\beta}(\omega)$ is  positive-semidefinite for an arbitrary $\omega$, Eqs.~(\ref{EqUni1}) and (\ref{EqUniS}) are of the first standard GKLS form.

Equations (\ref{EqUni1}) and (\ref{EqUniS}) are different expressions of the \textit{unified quantum master equation of weak-coupling limit type}.  A simple algorithm of its construction in the Schr\"odinger picture is as follows: 

(i) The dissipator $\mathcal D$ is constructed as if the system Hamiltonian was $H_S^{(0)}$, with the secular approximation with respect to $H_S^{(0)}$.

(ii) The Lamb-shift Hamiltonian $H_{\rm LS}$ is as for the Redfield equation with the  secular approximation with respect to $H_S^{(0)}$ [compare Eqs.~(\ref{EqHLSRedf}) and~(\ref{EqHLS})].

If we want to describe a concrete physical system, a question about the value of $\lambda$ arises. Formally, in order to apply the proposed analysis to a concrete physical system, we should express $H_S-H_S^{(0)}$ and the spectral density of the bath (indicating the strength of the system-bath coupling) as products of a small dimensionless parameter $\lambda^2$ and energy quantities of the same order as the zeroth-order system energies $\varepsilon_k^{(0)}$. In this case, $\lambda^2$ is the actual ratio of the scale of the small parameters to the scale of the large parameters (both are of energy dimensionality). So, $\lambda$ is defined up to an order of magnitude.

Moreover, $\lambda$ can be treated to be incorporated into the Hamiltonian (as it is often assumed in the physical literature). Indeed, if we denote $\lambda^2\delta H_S=\delta H'_S$ and $\lambda H_I=H'_I$, then $\lambda^2 H_{\rm LS}$ and $\lambda^2\mathcal D$ in Eq.~(\ref{EqUniS}) can be substituted by $H'_{\rm LS}$ and $\mathcal D'$, where the latter expressions are derived for the interaction Hamiltonian $H'_I$. The system Hamiltonian is then expressed as $H_S=H_S^{(0)}+\delta H'_S$. So, in practice,
for a given system, we should express its Hamiltonian as a sum of a reference part $H_S^{(0)}$, where all nearly degenerate Bohr frequencies are exactly degenerate, and a small perturbation $\delta H'_S$, and then apply Eq.~(\ref{EqUniS}) (formally, with $\lambda=1$ since $\lambda$ is implicitly incorporated into the Hamiltonian now). An explicit separation of a small factor $\lambda$ was required only in a formal derivation.

Let us summarize the physical conditions of validity
of master equation (\ref{EqUniS}): 
%\begin{enumerate}[(i)]

%\item 

(i) The usual weak system-bath coupling condition \cite{AL,BP,RH,MK,TrushLOJM}.  Roughly, it can be expressed as 
%\begin{equation}\label{EqWeakCoupCond}
$|\Gamma(\omega)|\ll \Omega$
%\end{equation}
 for all Bohr frequencies $\omega$, where $\Omega$ is a characteristic decay rate of the bath correlation functions (\ref{EqCorr}). For the Drude--Lorentz spectral density, $\Omega$ is an explicit parameter, see below. This condition means that the dissipative dynamics of the system is much slower than the bath relaxation. As a consequence, the system-bath state is always close to $\rho(t)\otimes\sigma_B$, which leads to a Markovian dissipative dynamics for the system.

%!!! 
(ii) Secular approximation with respect to $H^{(0)})_S$, i.e., oscillations $e^{i(\overline\omega'-\overline\omega)t}$ are much faster than the dissipative dynamics. Formally, $|\Gamma(\overline\omega)|\ll|\overline\omega'-\overline\omega|$ for all different $\overline\omega$ and $\overline\omega'$.

(iii) The functions $\Gamma_{\alpha\beta}(\omega)$ do not change significantly within the clusters of Bohr frequencies emerging due to the perturbative part $\delta H'_S$ of the system Hamiltonian. Formally, 
\begin{eqnarray*}
|\Re[\Gamma'_{\alpha\beta}(\overline\omega)]|\Delta\omega&\ll& |\Re[\Gamma_{\alpha\beta}(\overline\omega)]|,\\
|\Im[\Gamma'_{\alpha\beta}(\overline\omega)]|\Delta\omega&\ll& |\Im[\Gamma_{\alpha\beta}(\overline\omega)]|,
\end{eqnarray*}
where 
\begin{equation*}
\Delta\omega=\max_{\omega\in\mathcal F_{\overline\omega}}
|\omega-\overline\omega|.
\end{equation*}

%\end{enumerate}

\subsection{Comparison with the secular master equation}

Let us describe explicitly the terms neglected in the secular (Davies) master equation and taken into account in the presented unified master equation. We will use the common terms ``populations'' and ``coherences'' for, respectively, the diagonal and off-diagonal elements of the density matrix in some energy eigenbasis (not unique if some levels are degenerate). If $H_S\ket{e_i}=\varepsilon_i\ket{e_i}$ and $H_S\ket{e_j}=\varepsilon_j\ket{e_j}$, we say that the coherence $\braket{e_i|\rho|e_j}$ corresponds to the Bohr frequency $\varepsilon_j-\varepsilon_i$.

The secular master equation describes (i) transfer between populations, (ii) decay of coherences, (iii) transfer between coherences corresponding to equal Bohr frequencies, and (iv) transfer between populations and coherences corresponding to the zero Bohr frequency (i.e., coherences inside the eigensubspaces of $H_S$). 

The unified master equation describes the same processes with the following corrections: (iii$'$) transfer between coherences corresponding to close Bohr frequencies and (iv$'$) transfer between populations and coherences corresponding to Bohr frequencies which are equal or close to zero (i.e., coherences inside the eigensubspaces of $H_S^{(0)}$). 

This is illustrated on Fig.~\ref{Fig1} for the example given below.

\subsection{Comparison with other non-secular GKLS master equations}

Equations similar to the unified quantum master equation (but not exactly the same) appeared in the framework of the partial secular approximation \cite{PtaEsp,Bondar,NathanRudner,HartmannStrunz,CSP}. Let us compare our equation with the similar ones. Grouping the Bohr frequencies and taking the function $\gamma$ in the centers of the clusters makes the equation simpler in comparison with those in Refs.~\cite{PtaEsp,Bondar,NathanRudner}. Also, due to this, the obtained equation has the desired properties important for thermodynamics; see below. 

In contrast to one of equations in Ref.~\cite{HartmannStrunz} (derived in more detail in Ref.~\cite{CSP}), which also adopts the clustering of the Bohr frequencies and $\Gamma$ taken in the centers of the clusters, the free dynamics in Eq.~(\ref{EqUni1}) (manifested in the exponents) is defined by the original Hamiltonian $H_S$ (not $H^{(0)}_S$). Also, the functions $S_{\alpha\beta}$ in the Lamb-shift Hamiltonian have their original arguments.

\subsection{Particular cases. Refined Lamb-shift Hamiltonian}
If $\delta H_S=0$, then the unified master equation is reduced to the  Davies master equation. In contrast, the case $H_S^{(0)}=0$ (i.e., all Bohr frequencies are small) is known to be equivalent to the so-called singular coupling limit \cite{AL,BP,RH,Palmer,Spohn1980,AccFriLu}. In this case, the unified master equation  coincides with the master equation obtained in this limit, but with the refined Lamb-shift Hamiltonian. 

We can simplify the Lamb-shift Hamiltonian $H_{\rm LS}$ (\ref{EqHLS}) if we replace $S_{\alpha\beta}(\omega,\omega')$ by $S_{\alpha\beta}(\overline\omega,\overline\omega)$ (i.e., perform the limit $\lambda\to0$ in the arguments of the functions $S_{\alpha\beta}$, analogously to $\gamma_{\alpha\beta}$). In other words, both the dissipator and the Lamb-shift Hamiltonian would be constructed as if the system Hamiltonian was $H_S^{(0)}$. If $H_S^{(0)}=0$, the corresponding master equation is the well-known master equation for the singular coupling limit. But in some cases (see Appendix~\ref{SecLS}), solutions of this master equation  significantly deviate from the exact dynamics, while the  proposed master equation with the refined Lamb-shift Hamiltonian gives good results. So, we have obtained an improved version of the singular coupling limit master equation.

From the other side, we could keep the Lamb-shift Hamiltonian from the Redfield equation $\overline H_{\rm LS}$ (i.e., without even partial secular approximation). The equation would be still of the GKLS form, but without the desired thermodynamic properties.

\subsection{Properties} 

The Davies generator is well known to be covariant with respect to the unitary group $e^{iH_St}$ \cite{SpohnLeb}. This simplifies the structure of the dynamics and the steady states. Moreover, this property is related to the total (system and bath) energy conservation (hence, to the first law of thermodynamics) and for the resource theory of coherence \cite{Lostaglio,MarvianSpek, DannKosloff}. The unified generator (\ref{EqUniS}) shares this property, but with respect to the unitary group $e^{iH_S^{(0)}t}$:
\begin{equation}\label{EqCov}
e^{-iH_S^{(0)}t}(\mathcal L\rho)e^{iH_S^{(0)}t}
=
\mathcal L\big(e^{-iH_S^{(0)}t}\rho\, e^{iH_S^{(0)}t}\big)
\end{equation}
This relation is satisfied in view of Eq.~(\ref{EqAeigen}) and since $H_S^{(0)}$ commutes with both $H_S$ and $H_{\rm LS}$  ($H_S=\sum_kP_k^{(0)}H_SP_k^{(0)}$ and the same is true for $H_{\rm LS}$). Note that $H_{\rm LS}$ commutes with $H_S^{(0)}$, but, in general, not with $H_S$.

If the bath is thermal with the inverse temperature $\beta$ (not to be confused with the subindex), then the Kubo--Martin--Schwinger (KMS) condition 
$\gamma_{\alpha\beta}(-\overline\omega)=e^{-\beta\overline\omega}\gamma_{\beta\alpha}(\overline\omega)$ guarantees the stationarity of the thermal (Gibbs) state with respect to the bare system Hamiltonian $H_S^{(0)}$: $\mathcal L\rho_{\beta}=0$, where
$\rho_{\beta}=e^{-\beta H_S^{(0)}}/\Tr e^{-\beta H_S^{(0)}}$. The same properties guarantee that the quantum dynamical semigroup $e^{\mathcal Lt}$ satisfy the detailed balance property \cite{AL,SpohnLeb,Gorini} with respect to $\rho_\beta$.

\begin{remark}
Note that the true system steady state is expected to be not the Gibbs state with respect to $H_S$ (i.e., proportional to $e^{-\beta H_S}$), but the so-called mean force Gibbs state, i.e., proportional to $\Tr_Be^{-\beta(H_S+\lambda H_I)}$. In the zeroth order with respect to $\lambda$, it coincides with $\rho_{\beta}$ (the Gibbs state with respect to $H_S^{(0)}$). However, the first nontrivial correction (proportional to $\lambda^2$) does not coincide with the Gibbs state with respect to $H_S$, but takes into account system-bath steady-state correlations \cite{CresserAnders}.
\end{remark}

Let the system weakly interact with several baths with the inverse temperatures $\beta_n$ and the corresponding generators $\mathcal L_n$, so that $\mathcal L=-i[H_S,\,\cdot\,]+\lambda^2\sum_n\mathcal L_n$. We can consider the quantity of entropy production according to the general formalism \cite{SpohnLeb}. It is a sum of the increase of the von Neumann entropy of the system $S(\rho)=-\Tr\rho\ln\rho$ and the entropy flows from the system to the baths:
\begin{eqnarray*}
\frac{d}{dt}S(\rho(t))-\lambda^2\sum_n\beta_n\Tr\big\{H_S\,\mathcal L_n[\rho(t)]\big\}.
\end{eqnarray*}
Up to terms of the order $O(\lambda^4)$ (i.e., of a higher order with respect to the second-order master equation), $H_S$ here can be substituted by $H_S^{(0)}$. Then the expression is non-negative according to the analysis in Ref.~\cite{SpohnLeb} since $\mathcal L_n\rho_{\beta_n}=0$. We cannot calculate the fourth-order contribution to the entropy production using the second-order master equation. For the same reason, all other thermodynamic properties are also satisfied with respect to the zeroth-order system Hamiltonian $H_S^{(0)}$. In other words, resolution of small energy differences in the weak-coupling regime requires higher order corrections to the dissipator.

\section{Example: Two weakly interacting qubits}

Consider a system of two weakly interacting qubits with the Hamiltonian 
\begin{equation}\label{EqHS}
H_S=E_1\sigma_z^{(1)}+E_2\sigma_z^{(2)}+J\sigma_x^{(1)}\sigma_x^{(2)},
\end{equation}
where $\sigma_z=\ket1\bra1-\ket0\bra0$ and $\sigma_x=\ket1\bra0+\ket0\bra1$ are the Pauli matrices, the superindex denotes a qubit, $E_1\geq E_2>0$. Let qubits interact with three bosonic thermal baths (with different temperatures). The interaction Hamiltonian is $H_I=H_{I,0}+H_{I,1}+H_{I,2}$, $H_{I,j}=A_j\otimes B_j$. Here $A_j=\kappa_{jx}\sigma_x^{(j)}+
\kappa_{jz}\sigma_z^{(j)}$, where $\kappa_{jx}$ and $\kappa_{jz}$ are real numbers and, for the unified notations, we put $\sigma_{z,x}^{(0)}=\sigma_{z,x}^{(1)}+\sigma_{z,x}^{(2)}$. Further, $B_j=\int\,dk\,[\overline g_j(k)a_j(k)+g_j(k) a_j^\dag(k)]$, where $a_j(k)$ [$a^\dag_j(k)$] is an annihilation [creation] operator for the $k$th mode of the $j$th bath and $g_j(k)$ are complex-valued functions. So, bath 0 is a common bath interacting with both qubits and baths 1 and 2 are individual baths for the corresponding qubits. The model has been taken from Ref.~\cite{CattaneoLocGlob}. In general, a two-qubit system is nontrivial (in particular, has nontrivial thermodynamic properties) and often used as a benchmark for comparison of various descriptions of open quantum system dynamics \cite{CresserFacer,LevyKosloff,Decordi,TrushVol,Hofer}.

\begin{figure}[t]
\begin{centering}
\includegraphics[scale=1]{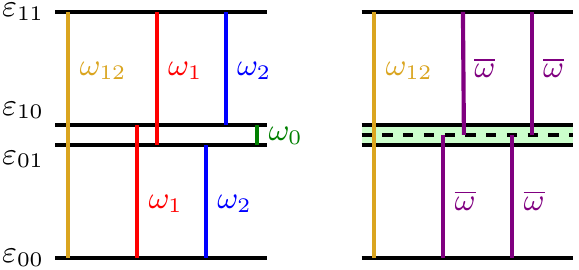}
\vskip -2mm
\caption
{\small
Scheme of energy levels and Bohr frequencies of two weakly interacting qubits. Left: The secular quantum master equation treats all distinct Bohr frequencies as well separated. Right: Nearly degenerate Bohr frequencies constitute clusters. Namely, the frequencies $\omega_1$ and $\omega_2$ constitute a cluster with the center at $\overline\omega=(\omega_1+\omega_2)/2$ and the frequencies 0 and $\pm\omega_0$ constitute a cluster with the center at 0.
}
\label{Fig1}
\end{centering}
\end{figure}

The eigenvalues and eigenvectors of $H_S$ are as follows:
\begin{eqnarray*}
\varepsilon_{11}&=&+\sqrt{E_{12}^2+J^2},\quad
\ket{e_{11}}=\cos\theta\ket{11}+\sin\theta\ket{00},\\
\varepsilon_{00}&=&-\sqrt{E_{12}^2+J^2},\quad
\ket{e_{00}}=\cos\theta\ket{00}-\sin\theta\ket{11},\\
\varepsilon_{10}&=&+\sqrt{\Delta E^2+J^2},\quad
\ket{e_{10}}=\cos\varphi\ket{10}+\sin\varphi\ket{01},\\
\varepsilon_{01}&=&-\sqrt{\Delta E^2+J^2},\quad
\ket{e_{01}}=\cos\varphi\ket{01}-\sin\varphi\ket{10},
\end{eqnarray*}
where $E_{12}=E_1+E_2$, $\Delta E=E_1-E_2$, $\theta=\frac12\arctan\frac{J}{E_{12}}$, and $\varphi=\frac12\arctan\frac{J}{\Delta E}$. The Bohr frequencies are shown on Fig.~\ref{Fig1}. We consider the case of large $E_1$ and $E_2$, but small $\Delta E$ and $J$, so that the energy levels 01 and 10 are almost degenerate and close to zero. Also the Bohr frequencies $\omega_1$ and $\omega_2$ almost coincide. 

So, we choose the following decomposition (\ref{EqHSdecompose}): 
\begin{equation*}
\begin{split}
H_S&=\varepsilon_{11}P_{11}
+\varepsilon_{00}P_{00}
+(\varepsilon_{10}P_{10}
+\varepsilon_{01}P_{01})\\
&=\varepsilon_{11}P_{11}
+\varepsilon_{00}P_{00}
+\lambda^2(\varepsilon'_{10}P_{10}
+\varepsilon'_{01}P_{01}),
\end{split}
\end{equation*}
where $P_{ab}=\ket{e_{ab}}\bra{e_{ab}}$.
Here $\lambda^2$ is a ratio of a characteristic large energy (i.e., $\varepsilon_{11}$) to a characteristic small energy (e.g., $\varepsilon_{10}$ or spectral densities of the baths), so that $\varepsilon'_{10,01}=\varepsilon_{10,01}/\lambda^2$ are of the order of $\varepsilon_{11}$ and $\varepsilon_{00}$.

With such decomposition, we have five clusters of Bohr frequencies: the clusters $\mathcal F_{+\overline\omega}=\{\omega_1,\omega_2\}$ and $\mathcal F_{-\overline\omega}=\{-\omega_1,-\omega_2\}$ with the centers at $\pm\overline{\omega}=\pm\varepsilon_{11}$, the cluster $\mathcal F_{\overline 0}=\{\pm\omega_0,0\}$ with the center at 0, and the clusters with single elements $\mathcal F_{\omega_{12}}=\{\omega_{12}\}$ and $\mathcal F_{-\omega_{12}}=\{-\omega_{12}\}$. 

Now we have all information to construct the unified maser equation; see Appendix~\ref{SecDetails} for details.

%\section{Calculations} 

\begin{figure}[t]
\begin{centering}
\includegraphics[width=\columnwidth]{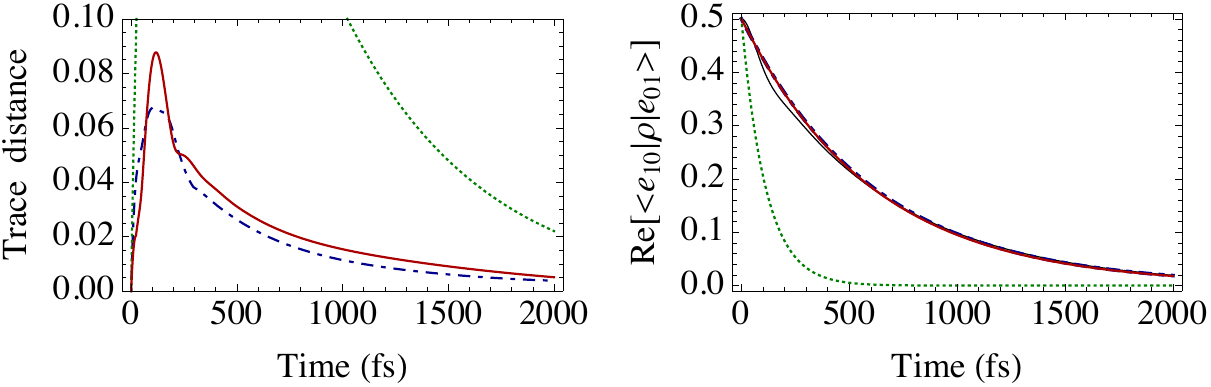}
\vskip -4mm
\caption
{\small
Comparison of calculations of the dynamics of the two-qubit system according to the Redfield equation (blue dash-dotted line), secular (Davies) master equation (green dotted line), and the unified approach (red solid line). Left: Trace distance to the numerically exact solution  of the HEOM. Right: The dynamics of the coherence that gives the main contribution to the error for the secular master equation. The solution of the HEOM is shown by the black solid line. The parameters are given in the  text.
}
\label{Fig2}
\end{centering}
\end{figure}

An example of calculations is presented in Fig.~\ref{Fig2}. We have chosen the same Drude--Lorentz spectral density for all three baths, 
\begin{equation}\label{EqDrude}
\mathcal J_j(\omega)\equiv\int|g_j(k)|^2\delta(\omega(k)-\omega)\,dk=
\frac{2\eta\Omega\omega}{\pi(\omega^2+\Omega^2)},
\end{equation}
with $\eta=1~\rm{cm}^{-1}$ and $\Omega^{-1}=100~\rm{fs}$ ($\Omega\approx53.08~\rm{cm}^{-1}$). The temperatures of the baths are $T_1=300~\rm{K}$, $T_2=400~\rm{K}$, and $T_0=350~\rm{K}$. The parameters of the system Hamiltonian are as follows: $E_1=E_2=50~\rm{cm}^{-1}$, $J=2~\rm{cm}^{-1}$. These parameters of the baths and the system Hamiltonian are mainly taken from Ref.~\cite{IFl} as model parameters for excitation energy transfer in a molecular dimer. Let us specify the interaction Hamiltonian: $\kappa_{jx}=0$, and $\kappa_{jz}=1$ for $j=1,2$, $\kappa_{0x}=1$ and $\kappa_{0z}=0$. The initial state is  $\rho(0)=\ket{01}\bra{01}$. %!!!

%!!!
In Fig.~\ref{Fig2}, we compare solutions according to different  master equations with the numerically exact nonperturbative method of the hierarchical equation of motion (HEOM) in the high-temperature approximation, according to Ref.~\cite{IFl}. We see the breakdown of the secular master equation, while the unified master equation correctly describes the dynamics.

It is worthwhile to recall that the HEOM is a numerically exact but  computationally expensive method (in terms of both time and memory),  especially for the low-temperature case. Quantum master equations are much simpler to solve numerically. Also, a relatively simple structure of the master equation gives more insights into mechanisms of various quantum dynamical phenomena. A two-qubit system interacting with high-temperature reservoirs was chosen since it is a simple (but nontrivial) model which can be solved also by the HEOM; i.e., approximate descriptions can be compared with the numerically exact one.

\section{Discussion}

\subsection{Global vs local approach} 

The local and global approaches to open quantum dynamics are largely debated in the literature \cite{BreuerLoc,RHP,Hofer,Gonzalez,Correa,LevyKosloff,TrushVol,Decordi,
CattaneoLocGlob,TruBashJETP,TruBashScr}. On the one hand, the unified approach is global, i.e., adopts the eigenvalues and eigenprojectors of the whole system Hamiltonian $H_S$ rather than of the Hamiltonian of noninteracting sites (corresponding to $J=0$ in our example). On the other hand, in our example, $J\ll E_1+E_2$, so, we can  neglect $\theta$ and approximately put $\ket{e_{00}}=\ket{00}$ and $\ket{e_{11}}=\ket{11}$. This corresponds to the rotating-wave approximation in the system Hamiltonian. Thus, $H_S^{(0)}$ corresponds to  $J=0$ and leads to a local dissipator. This is exactly the case where the local approach is shown to be better than the secular global one \cite{BreuerLoc,RHP,Hofer,Gonzalez,CattaneoLocGlob,Correa}. However, the same factor $\gamma_j(\overline\omega)$ in the dissipators of the two local baths instead of different factors $\gamma_j(2E_1)$ and $\gamma_j(2E_2)$ in the ``naive'' local approach make the obtained local equation consistent with the second law of thermodynamics. 

Thus, if the intersite couplings are small, the local approach can be used, but the arguments of the dissipation coefficients $\gamma$ should be the same for the sites with close local energies. If the difference between the local energies ($\Delta E$ in our example) is large such that the secular approximation can be used, but the intersite couplings are still small, then the local approach can be justified as an approximation to the secular or unified (i.e., global) quantum master equation \cite{TrushVol}.

\subsection{Unified vs Redfield master equation}

Our numerical results (see Fig.~\ref{Fig2} and also Fig.~\ref{Fig3} in Appendix~\ref{SecLS}) suggest that the Redfield equation is, in most cases, more precise that the unified GKLS master equation. This agrees with the results of Ref.~\cite{HartmannStrunz}. So, the choice between the Redfield equation and the unified GKLS master equation depends on our purposes. If our priority is the numerical precision and we do not care the possible small violation of positivity, then the Redfield equation is preferable. If we want to have an equation with good theoretical properties, with the absence of nonphysical predictions, and with a reasonable error, then  the proposed unified GKLS master equation of weak-coupling limit type is a good candidate. 

Also, a ``hybrid'' variant can be used to use the unified GKLS master equation for the initial short-time period (where the Redfield equation can violate positivity) and then to switch to the Redfield equation. Other concatenation schemes for open quantum systems where different descriptions are used for the initial short-time period and for further time scales are considered in Refs.~\cite{ChengSilbey,Tere}.

\subsection{Arbitrary scaling of the Bohr frequency spacing}

Instead of Eq.~(\ref{EqHSdecompose}), an arbitrary scaling of the Bohr frequency spacing can be considered:
\begin{equation}
H_S=H_S^{(0)}+\lambda^\nu \delta H_S,\quad\nu>0.
\end{equation}
Rigorous results for particular models were obtained in Refs.~\cite{DaviesAtomRad,MerkliSong,MerkliSongBerman}. However, as these results suggest, a kind of a dynamical phase transition occurs exactly in the case $\nu=2$. This can be understood also in terms of the present analysis. Namely, $\nu<2$ means that the oscillations $e^{i\lambda^\nu(\omega'-\omega)\tau/\lambda^2}$ in Eq.~(\ref{EqRedfieldGKLS}) are much faster than the dissipative dynamics. Hence, the full secular approximation can be applied, which leads to the Davies master equation. 

%!!! %!!! 
If, on the contrary, $\nu>2$, then these oscillations are much slower than the dissipative dynamics. This leads to emergence of the two time scales described in Refs.~\cite{DaviesAtomRad,MerkliSong,MerkliSongBerman}: a relaxation to a manifold of quasistationary states (stationary if $\delta H_S=0$) and a slower relaxation to a final stationary state. Formal derivation for this scaling according to Sec.~\ref{SecDeriv} again leads to the unified master equation. Namely, the secular approximation with respect to $H^{(0)}_S$ and the limit $\lambda\to0$ in the arguments of $\gamma_{\alpha\beta}$ are still valid. So, the unified master equation can be applied irrespective of whether some oscillations occur on the same time scale as the dissipative dynamics or on a larger scale.

\section{Conclusions} 

The unified approach allows one to derive the correct quantum master equation for a specific physical systems (for a specific structure of energy levels) in a rigorous and systematic way. The unified master equation has the GKLS form (hence, preserves positivity) and all the desired properties important for thermodynamics: stationarity of the Gibbs state, the detailed balance condition, the covariance law related to the first law of thermodynamics, and non-negativity of the entropy production (the second law of thermodynamics). Thus, the unified quantum master equation can be used in a wide range of physical applications.

\begin{acknowledgments}
I am grateful to Marco Cattaneo, Dariusz Chru\'{s}ci\'{n}ski, Luis A. Correa, Camille Lombard Latune, \'Angel Rivas, Thomas Schulte-Herbr\"uggen, Ilya Sinayskiy, and Alexander Teretenkov for fruitful discussions and some bibliographic references. This work was funded by Russian Federation represented by the Ministry of Science and Higher Education (Grant No. 075-15-2020-788).
\end{acknowledgments}

\bigskip
\appendix

\section{Rigorous result}\label{SecRig}

We consider a quantum system specified by an either finite- or infinite-dimensional Hilbert space $\mathcal H_S$ and a Hamiltonian of the form 
\begin{equation}\label{EqHSdecomposeCopy}
H_S\equiv H_S^{(\lambda)}=H_S^{(0)}+\lambda^2\delta H_S,
\end{equation}
where the operators $H_S^{(0)}$ and $\delta H_S$ are commuting self-adjoint operators with purely discrete spectra. Hence, the spectrum of the operator $H_S^{(\lambda)}$ is also purely discrete. Also we assume that the operator $\delta H_S$ is bounded. 

Like in the Davies's paper \cite{Davies}, for simplicity, we consider the case when the bath is fermionic. Namely, the bath is described by a quasi-free representation of the canonical anticommutation relations (CAR) with an infinite number of degrees of freedom. Denote the corresponding Hilbert space by $\mathcal H_B$ and the single-particle Hilbert space by $\mathcal H_B^{(1)}$. For each $f\in\mathcal H_B^{(1)}$, there is a bounded operator $a(f)$ acting on $\mathcal H_B$ with an antilinear dependence on $f$ which satisfies the anticommutation relations
\begin{equation*}
\begin{split}
a(f)a^\dag(g)+a^\dag(g)a(f)&=\braket{f|g},\\
a(f)a(g)+a(g)a(f)&=0.
\end{split}
\end{equation*} 

The single-particle free evolution is given by $f_t=e^{-ith}f$, where $h$ is the single-particle Hamiltonian (on $\mathcal H_B^{(1)}$). There is a Hamiltonian $H_B$ on $\mathcal H_B$ and a cyclic vector $\ket\Omega\in\mathcal H_B$ such that $H_B\ket{\Omega}=0$ and
\begin{equation*}
e^{iH_Bt}a(f)e^{-iH_Bt}=a(e^{iht}f).
\end{equation*}

If, for example, $\mathcal H_B^{(1)}=L^2(\mathbb R^d)$, then the anticommutation relations can be rewritten in terms of the operator-valued distributions:
\begin{equation*}
\begin{split}
a(k)a^\dag(k')+a^\dag(k')a(k)&=\delta(k-k'),\\
a(k)a(k')+a(k')a(k)&=0,
\end{split}
\end{equation*} 
$k,k'\in\mathbb R^d$.
The distribution and operator pictures are related to each other by the formula
\begin{equation*}
a(f)=\int_{\mathbb R^d} \overline{f(k)} a(k)\,dk.
\end{equation*}  
If, further, $hf(k)=\omega(k)f(k)$ for a real-valued function $\omega(k)$, then $f_t(k)=e^{-it\omega(k)}f(k)$ and
\begin{equation*}%\label{EqHB}
H_B=\int_{\mathbb R^d}\omega(k)a^\dag(k)a(k)\,dk.
\end{equation*}

Writing $\langle\cdot\rangle$ for the expectation with respect to $\ket\Omega$, we have
\begin{equation}\label{EqMultiCorr}
\langle a^\dag(g_m)\cdots a^\dag(g_1)a(f_1)\cdots a(f_n)\rangle
=
\delta_{nm}\,{\rm det}\{\braket{f_i|R|g_j}\}
\end{equation}
(the quasi-free state property),
where $R$ is the defining operator on $\mathcal H_B^{(1)}$. Since $H_B\ket\Omega=0$, we have $e^{iht}Re^{-iht}=R$. For the equilibrium state at the inverse temperature $\beta$ and the chemical potential $\mu$, 
\begin{equation*}
R=(e^{\beta(h-\mu)}+1)^{-1}.
\end{equation*}

The full Hamiltonian is given by
\begin{equation*}
H^{(\lambda)}=H_S^{(\lambda)}+H_B+\lambda H_I,
\end{equation*}
where the interaction Hamiltonian $H_I$ is given by a finite sum $H_I=\sum_\alpha A_\alpha\otimes B_\alpha$ with $A_\alpha$ and $B_\alpha$ being bounded operators in $\mathcal H_S$ and $\mathcal H_B$, respectively. We assume that each $B_\alpha$ is a linear combination of $a(g_\alpha)$ and $a^\dag(g_\alpha)$ for some $g_\alpha\in\mathcal H_B^{(1)}$. In particular, $\langle B_\alpha\rangle=0$. Denote $C_{\alpha\beta}(t)=\langle e^{iH_Bt} B_\alpha^\dag e^{-iH_Bt} B_\beta\rangle$
the correlation functions and let
\begin{equation}\label{EqCint}
\int_0^\infty |C_{\alpha\beta}(t)|(1+t)^\delta\,dt<\infty
\end{equation}
for some $\delta>0$. 

The evolution of the open quantum system is given by
\begin{equation*}
\mathcal T^{(\lambda)}_t\rho=
\Tr_B
\big\{
e^{-iH^{(\lambda)}t}
(\rho\otimes\sigma_B)
e^{iH^{(\lambda)}t}
\big\}
\end{equation*}
for an arbitrary trace-class operator $\rho$ on $\mathcal H_S$, where $\sigma_B=\ket\Omega\bra\Omega$. 

\begin{remark}
Let us stress that $\ket\Omega$ is the vacuum vector in a Fock space only in the case of the zero temperature. Otherwise, this is a cyclic vector in the representation of the CAR algebra. In the physical literature, the notation $e^{-\beta H_B}/\Tr e^{-\beta H_B}$ is used for a thermal state instead of $\ket\Omega\bra\Omega$, but, strictly speaking, the former is not a genuine density operator because the trace of $e^{-\beta H_B}$ is ill defined in the case of an infinite number of degrees of freedom. 

%!!!
As a remedy, in the physical literature, one often considers a finite number $N$ of the bath modes and then tends this number to infinity. But, strictly speaking, this may cause issues with the order of the limit $N\to\infty$ and the Bogolyubov--van Hove limit $\lambda\to0$, $t\to\infty$, $\lambda^2t=\tau={\rm const}$.

So, in the rigorous derivation, an infinite number of degrees of freedom is considered from the beginning, and a thermal state can be understood only in a ``generalized'' sense: as a functional on the CAR algebra. The state represented by a cyclic vector $\ket\Omega$ with property (\ref{EqMultiCorr}) is a formalization of this situation. Thus, physically, $\sigma_B$ is an arbitrary stationary state of the bath with property (\ref{EqMultiCorr}). In particular, it can be associated with the thermal state with an arbitrary temperature.
\end{remark}

\begin{theorem}
Under the described assumptions, 
\begin{equation}\label{EqTh}
\lim_{\lambda\to0}\sup_{0\leq\lambda^2t\leq\tau_1}
\|\mathcal T^{(\lambda)}_t\rho-
e^{(-i[H_S^{(\lambda)}+\lambda^2 H_{\rm LS},\,\cdot\,]+\lambda^2\mathcal D)t}\rho\|=0
\end{equation}
for any $\tau_1>0$ and a trace-class operator $\rho$ on $\mathcal H_S$. 
\end{theorem}
Here $\mathcal D$ and $H_{\rm LS}\equiv H_{\rm LS}^{(\lambda)}$  are given in Eqs.~(\ref{EqD}) and~(\ref{EqHLS}) and $\|\cdot\|$ denotes the trace norm.

\begin{proof}
Let us introduce the operators $\mathcal L_0^{(\lambda)}=[H_S^{(\lambda)}+H_B,\,\cdot\,]$, $\mathcal L_I=[H_I,\,\cdot\,]$,  and $\mathcal L_I^{(\lambda)}(s)=[H_I^{(\lambda)}(s),\,\cdot\,]$
acting on the Banach space $\mathcal B$ of the trace-class operators on $\mathcal H_S\otimes\mathcal H_B$, and the operators $\mathcal L_S^{(\lambda)}=[H_S^{(\lambda)},\,\cdot\,]$ and $\delta\mathcal L_S=[\delta H_S,\,\cdot\,]$ acting on the Banach space $\mathcal B_S$ of the trace-class operators on $\mathcal H_S$. Here $H^{(\lambda)}_I(s)=e^{i(H_S^{(\lambda)}+H_B)s}H_Ie^{-i(H_S^{(\lambda)}+H_B)s}$.
%!!!

According to Theorems~1.2 and~1.3  of Ref.~\cite{Davies2}, Theorems~3.1--3.5 of Ref.~\cite{Davies} (with minor modifications noted in Ref.~\cite{Palmer}), and assumptions of our theorem, 
\begin{equation}\label{EqDaviesThK}
\lim_{\lambda\to0}\sup_{0\leq \lambda^2 t\leq\tau_1}
\big\|\,\mathcal T_t^{(\lambda)}-e^{(-i\mathcal L_S^{(\lambda)}+
\lambda^2\mathcal K^{(\lambda)})t}\,\big\|=0,
\end{equation}
where
\begin{equation}\label{EqK}
\mathcal K^{(\lambda)}\rho=-\int_0^\infty 
\Tr_B
\big[
\mathcal L_I^{(\lambda)}(s)\mathcal L_I
(\rho\otimes\sigma_B)
\big]
\,ds
\end{equation}
is an operator on $\mathcal B_S$.

The norm of the integrand in Eq.~(\ref{EqK}) is upper bounded by the integrable function 
$$\sum_{\alpha,\beta}4\|A_\alpha A_\beta\|\cdot|C_{\alpha\beta}(s)|\cdot\|\rho\|$$
independent from $\lambda$ 
(the factor 4 comes from the two commutators in $\mathcal L_I$ and $\mathcal L_I^{(\lambda)}$ resulting in four terms). Since $H_S^{(0)}$ and $\delta H_S$ commute, $e^{it\mathcal L_0^{(\lambda)}}=e^{it\mathcal L_0^{(0)}}e^{i\lambda^2t\delta\mathcal L_S}$. Since $\delta H_S$ is bounded, $\|e^{i\lambda^2t\delta\mathcal L_S}-1\|\to0$ as $\lambda\to0$. So, by the Lebesgue's dominated convergence theorem, it is easy to show that $\|\mathcal K^{(\lambda)}-\mathcal K^{(0)}\|\to0$ as $\lambda\to0$. Then (the proof is completely analogous to that of Theorem~1.2 of Ref.~\cite{Davies2}),
\begin{equation}\label{EqDaviesThKK}
\lim_{\lambda\to0}\sup_{0\leq \lambda^2 t\leq\tau_1}
\big\|\,
e^{(-i\mathcal L_S^{(\lambda)}+
\lambda^2\mathcal K^{(\lambda)})t}-
e^{(-i\mathcal L_S^{(\lambda)}+
\lambda^2\mathcal K^{(0)})t}\,\big\|=0.
\end{equation}

It is interesting to note that the generator $-i\mathcal L_S^{(\lambda)}+
\lambda^2\mathcal K^{(\lambda)}$ differs from the usual Redfield generator, which is $-i\mathcal L_S^{(\lambda)}+
\lambda^2\mathcal R^{(\lambda)}$, where
\begin{equation*}
\mathcal R^{(\lambda)}=-\int_0^\infty 
\Tr_B
\big[
\mathcal L_I\mathcal L^{(\lambda)}_I(-s)
(\rho\otimes\sigma_B)
\big]
\,ds.
\end{equation*}
The generator $\mathcal K^{(\lambda)}$  is similar to the Redfield generator. In particular, it is also non-GKLS and does not preserve positivity. Its explicit form also can be described by Eqs.~(\ref{EqRedfieldGKLS}) and~(\ref{EqHLSRedf}) if we replace $\gamma(\omega,\omega')$ and $S(\omega,\omega')$ there by $\gamma(\omega',\omega)$ and $S(\omega',\omega)$.

The secular approximation with respect to the reference system Hamiltonian $H_S^{(0)}$ (which corresponds to a partial secular approximation with respect to the original system Hamiltonian $H_S^{(\lambda)}$) can be expressed as

\begin{equation}\label{EqSec}
\mathcal A^{\rm sec}=\lim_{T\to\infty}\frac1{2T}
\int_{-T}^T 
e^{i\mathcal L_S^{(0)}u}
\mathcal A
e^{-i\mathcal L_S^{(0)}u}\,du,
\end{equation}
for an arbitrary operator $\mathcal A$ on the Banach space $\mathcal B_S$. According to Theorem~1.4 of Ref.~\cite{Davies2},
\begin{equation}\label{EqDaviesThSec}
\lim_{\lambda\to0}
\sup_{0\leq\lambda^2t\leq\tau_1}
\|
e^{(-i\mathcal L_S^{(0)}+\lambda^2\mathcal A)t}\rho
-
e^{(-i\mathcal L_S^{(0)}+\lambda^2\mathcal A^{\rm sec})t}\rho
\|=0
\end{equation}
for all $\rho\in\mathcal B_S$. 

Obviously, the secular approximation applied to $\mathcal K^{(0)}$ and $\mathcal R^{(0)}$ gives the same result $-i\tilde{\mathcal L}_{\rm LS}+\mathcal D$, where $\mathcal D$ is as required [i.e., given by Eq.~(\ref{EqD})] and $\tilde{\mathcal L}_{\rm LS}=
[\tilde H_{\rm LS},\,\cdot\,]$. Here, in the notations of the main text,
\begin{equation}\label{EqHLSsec}
\tilde H_{\rm LS}=
\sum_{\overline\omega\in\mathcal F^{(0)}}
\sum_{\alpha,\beta}
S_{\alpha\beta}(\overline\omega)
A_{\alpha\overline\omega}^\dag A_{\beta\overline\omega},
\end{equation}
$S_{\alpha\beta}(\overline\omega)\equiv
S_{\alpha\beta}(\overline\omega,\overline\omega)$.

The term $\delta\mathcal L_S$ is already secular since $H^{(0)}_S$ and $\delta H_S$ commute. Hence, the secular approximation applied to $-i\delta\mathcal L_S+\mathcal K^{(0)}$ and $-i\delta\mathcal L_S+\mathcal R^{(0)}$ gives the same result. Since, due to Eq.~(\ref{EqDaviesThSec}), the corresponding semigroups are close to the same third semigroup, they are close to each other  in the same sense:
\begin{equation}\label{EqKD}
\lim_{\lambda\to0}
\sup_{0\leq\lambda^2t\leq\tau_1}
\|
e^{(-i\mathcal L_S^{(\lambda)}+\lambda^2\mathcal K^{(0)})t}\rho
-
e^{(-i\mathcal L_S^{(\lambda)}+\lambda^2\mathcal R^{(0)})t}\rho
\|=0
\end{equation}
for all $\rho\in\mathcal B_S$. 

The decomposition of $\mathcal R^{(\lambda)}=-i\overline{\mathcal L}_{\rm LS}^{(\lambda)}+\mathcal D^{(\lambda)}$ into the Lamb-shift Hamiltonian $\overline{\mathcal L}_{\rm LS}^{(\lambda)}=[\overline H_{\rm LS}^{(\lambda)},\,\cdot\,]$ and the dissipator $\mathcal D^{(\lambda)}$ [cf. Eq.~(\ref{EqRedfieldGKLS}), where $\overline H_{\rm LS}
\equiv \overline H_{\rm LS}^{(\lambda)}$ because $\overline H_{\rm LS}$ depends on the spectrum of $H_S^{(\lambda)}$] can be expressed in the general form as follows:

\begin{equation*}
\overline{\mathcal L}_{\rm LS}^{(\lambda)}\rho
=
\frac{1}{2i}
\int_0^\infty
\Tr_B[H_IH_I(-s)-H_I(-s)H_I,\rho\otimes\sigma_B]\,ds,
\end{equation*}
\begin{multline*}
\mathcal D^{(\lambda)}\rho
=\int_0^\infty
\Tr_B
\Big[
H_I(\rho\otimes\sigma_B)H_I^{(\lambda)}(-s)
\\+
H_I^{(\lambda)}(-s)(\rho\otimes\sigma_B)H_I
\\
-\frac12
\big\{
H_IH_I^{(\lambda)}(-s)
+H_I^{(\lambda)}(-s)H_I,\rho\otimes\sigma_B
\big\}
\Big]\,ds.
\end{multline*}

Analogously to the proof of Eq.~(\ref{EqDaviesThKK}), we can prove that
\begin{multline}\label{EqDaviesThDD}
\lim_{\lambda\to0}\sup_{0\leq \lambda^2 t\leq\tau_1}
\big\|\,
e^{(-i\mathcal L_S^{(\lambda)}
-i\lambda^2\overline{\mathcal L}_{\rm LS}^{(\lambda)}+
\lambda^2\mathcal D^{(\lambda)})t}
\\-
e^{(-i\mathcal L_S^{(\lambda)}
-i\lambda^2\overline{\mathcal L}_{\rm LS}^{(\lambda)}+
\lambda^2\mathcal D^{(0)})t}\,\big\|=0.
\end{multline}
Finally, due to Eq.~(\ref{EqDaviesThSec}) applied to $\mathcal A=-i\delta\mathcal L_S-i\overline{\mathcal L}^{(\lambda)}_{\rm LS}
+\mathcal D^{(0)}$, we obtain
\begin{multline}\label{EqDaviesThDDsec}
\lim_{\lambda\to0}\sup_{0\leq \lambda^2 t\leq\tau_1}
\big\|\,
e^{(-i\mathcal L_S^{(\lambda)}
-i\lambda^2\overline{\mathcal L}_{\rm LS}^{(\lambda)}+
\lambda^2\mathcal D^{(0)})t}\rho
\\-
e^{(-i\mathcal L_S^{(\lambda)}
-i\lambda^2 \mathcal L_{\rm LS}^{(\lambda)}+
\lambda^2\mathcal D)t}\rho
\,\big\|=0,
\end{multline}
for all $\rho\in\mathcal B_S$,
where $\mathcal L_{\rm LS}^{(\lambda)}=[H_{\rm LS}^{(\lambda)},\,\cdot\,]$ and $H_{\rm LS}^{(\lambda)}$ is obtained by dropping the nonsecular terms from $\overline H_{\rm LS}^{(\lambda)}$ and is given by Eq.~(\ref{EqHLS}). %!!!x2

A combination of Eqs.~(\ref{EqDaviesThK}), (\ref{EqDaviesThKK}), (\ref{EqKD}), (\ref{EqDaviesThDD}), and~(\ref{EqDaviesThDDsec}) gives the required Eq.~(\ref{EqTh}).
\end{proof}

\begin{remark}
The proof is a bit cumbersome because we wanted to have the Lamb-shift Hamiltonian from the Redfield equation (up to the secular approximation with respect to $H_S^{(0)}$). However, the Davies method gives another generator $-i\mathcal L_S^{(\lambda)}+\lambda^2\mathcal K^{(\lambda)}$, which is different from the Redfield one $-i\mathcal L_S^{(\lambda)}+\lambda^2\mathcal R^{(\lambda)}$. Both generators give the same result after the secular approximation and the Davies method does not say which one is more precise, but our numerical results suggest that the Redfield generator is more precise than the ``nonsecular Davies'' generator $-i\mathcal L_S^{(\lambda)}+\lambda^2\mathcal K^{(\lambda)}$. Also, our unified GKLS generator $-i\mathcal L_S^{(\lambda)}-i\lambda^2 \mathcal L_{\rm LS}^{(\lambda)}+
\lambda^2\mathcal D$ is more precise than the GKSL generator with the same $\mathcal D$, but the Lamb-shift Hamiltonian constructed from the operator $\mathcal K^{(\lambda)}$ instead of $\mathcal R^{(\lambda)}$ [i.e., with $S(\omega,\omega')$ in Eq.~(\ref{EqHLS}) replaced by $S(\omega',\omega)$].

%!!!x2
Also, mathematically, it would be more natural to deal with the generator with the same $\mathcal D$ and the simplified Lamb-shift Hamiltonian (\ref{EqHLSsec}), i.e., with the generator obtained by the application of the secular approximation to $\mathcal R^{(0)}$ or, equivalently, $\mathcal K^{(0)}$. Such possibility was discussed in the main text: In this case, both the dissipator and the Lamb-shift Hamiltonian are constructed as if the system Hamiltonian was $H_S^{(0)}$. Of course,
\begin{equation}\label{EqThSimp}
\lim_{\lambda\to0}\sup_{0\leq\lambda^2t\leq\tau_1}
\|\mathcal T^{(\lambda)}_t\rho-
e^{(-i[H_S^{(\lambda)}+\lambda^2 \tilde H_{\rm LS},\,\cdot\,]+
\lambda^2\mathcal D)t}\rho\|=0
\end{equation}
is also true as a consequence of Eqs.~(\ref{EqDaviesThK}), (\ref{EqDaviesThKK}), and~(\ref{EqDaviesThSec}) (so, the proof is easier). However, in Appendix~D, we give an example where such generator gives significant error, while the proposed generator with a refined Lamb-shift Hamiltonian gives good results.
\end{remark}

\begin{remark}
Expressions (\ref{EqTh}) and (\ref{EqThSimp}) assert that the quantum dynamical semigroups approach the exact reduced dynamics in the limit $\lambda\to0$ on arbitrarily long but finite time segments $[0,\tau_1]$ (in the rescaled time). This restriction is important if we study the long-time limit $\lim_{\tau\to\infty}\rho(\tau/\lambda^2)$. Recently \cite{MerkliRev}, for the standard weak-coupling limit (i.e., leading to the secular master equation), it was shown that the quantum dynamical semigroup approaches the exact reduced dynamics uniformly on the whole time half-line $\tau\in[0,\infty)$. Probably, the same result can be proved also for our limiting regime (\ref{EqHSdecomposeCopy}).

But, nevertheless, we know (see the same paper Ref.~\cite{MerkliRev} for a review of the rigorous results about this) that, if the bath is in thermal equilibrium and certain conditions are met, $\rho(t)$ tends to the state $\Tr_B e^{-\beta H^{(\lambda)}}/\Tr e^{-\beta H^{(\lambda)}}$ as $t\to\infty$ for an arbitrary initial state $\rho(0)$. As $\lambda\to0$, this stationary state tends to $\rho_{\beta}=e^{-\beta H_S^{(0)}}/\Tr e^{-\beta H_S^{(0)}}$. As we know from the main text, the unified master equation correctly predicts the stationarity of this state. Under the same conditions, this stationary state is unique.
\end{remark}

\section{Details of calculations in the example}\label{SecDetails}

The unified quantum master equation for the considered example has the form
\begin{equation*}
\dot\rho=-i[H_S,\rho]+
\sum_{j=0}^2\mathcal L_j\rho,
\end{equation*}
where $\mathcal L_j=-i[H_{\rm LS}^{(j)},\,\cdot\,]+\mathcal D_j$ is the generator corresponding to the interaction with the $j$th bath. Here $\mathcal D_j=\mathcal D_{j\omega_{12}}+\mathcal D_{j\overline\omega}+\mathcal D_{j\overline 0}$,
\begin{multline*}
\mathcal D_{j\omega}\rho=
\gamma_j(\omega)
\Big(
A_{j\omega}\rho A_{j\omega}^\dag
-
\frac12
\big\{A_{j\omega}^\dag A_{j\omega},\rho\big\}
\Big)\\
+\gamma_j(\omega)e^{-\beta_j\omega}
\Big(
A_{j\omega}^\dag\rho A_{j\omega}
-
\frac12
\big\{A_{j\omega} A_{j\omega}^\dag,\rho\big\}
\Big),
\end{multline*}
$\omega$ is either $\omega_{12}$ or $\overline\omega$, and
\begin{equation*}
\mathcal D_{j\overline0}\rho=
\gamma_j(0)
\Big(
A_{j\overline0}\rho A_{j\overline0}
-
\frac12
\big\{A_{j\overline0}^2,\rho\big\}
\Big).
\end{equation*}

The jump operator $A_{j\omega_{12}}$ corresponds to an individual Bohr frequency of $H_S$ and defined by Eq.~(\ref{EqAomega}): $A_{j\omega_{12}}=\kappa_{zj}P_{00}\sigma_z^{(j)}P_{11}$. Let us discuss the operators $A_{j\overline\omega}$ and $A_{j\overline0}$ corresponding to clusters of Bohr frequencies of $H_S$, or, equivalently, to individual Bohr frequencies of 
$$H_S^{(0)}=\varepsilon_{11}(P_{11}-P_{00})
+0\cdot(P_{01}+P_{10}).$$
As we see, the eigenprojectors of $H_S^{(0)}$ are $P_{+}^{(0)}=P_{11}$, $P_{-}^{(0)}=P_{00}$, and $P_0^{(0)}=P_{01}+P_{10}$.

According to Eq.~(\ref{EqAbaromega}), 
\begin{equation*}
\begin{split}
A_{j\overline\omega}&=
P_-^{(0)}A_jP_0^{(0)}+P_0^{(0)}A_jP_+^{(0)}
\\
&=P_{00}A_j(P_{01}+P_{10})+(P_{01}+P_{10})A_jP_{11},\\
&=\kappa_{xj}
[P_{00}\sigma_{x}^{(j)}(P_{01}+P_{10})
+
(P_{01}+P_{10})\sigma_{x}^{(j)}P_{11}],\\
A_{j\overline0}&=P_0^{(0)}A_jP_0^{(0)}
+P_+^{(0)}A_jP_+^{(0)}+P_-^{(0)}A_jP_-^{(0)}\\
&=\kappa_{zj}
[(P_{01}+P_{10})\sigma^{(j)}_z(P_{01}+P_{10})
\\&\qquad\:+P_{00}\sigma^{(j)}_zP_{00}+P_{11}\sigma^{(j)}_zP_{11}],
\end{split}
\end{equation*}
Alternatively, these operators can be calculated as sums of the operators corresponding to the individual Bohr frequencies of $H_S$ from the corresponding clusters: 
\begin{equation*}
\begin{split}
A_{j\overline\omega}&=A_{j\omega_1}+A_{j\omega_2},\\ A_{j\overline0}&=A_{j0}+A_{j\omega_0}+A_{j,-\omega_0},
\end{split}
\end{equation*}
where
\begin{equation*}
\begin{split}
&A_{j\omega_1}=\kappa_{xj}(P_{00}\sigma_x^{(j)}P_{10}
+P_{01}\sigma_x^{(j)}P_{11}),\\
&A_{j\omega_2}=\kappa_{xj}(P_{00}\sigma_x^{(j)}P_{01}
+P_{10}\sigma_x^{(j)}P_{11}),\\
&A_{j0}=\sum_{\mu=00,01,10,11}
P_\mu A_jP_\mu,\\
&A_{j\omega_0}=A_{j,-\omega_0}^\dag=P_{01}A_jP_{10}.
\end{split}
\end{equation*}

Further,
\begin{multline*}
H_{\rm LS}^{(j)}=S(\omega_{12})
A_{j\omega_{12}}^\dag A_{j\omega_{12}}+
S(-\omega_{12})
A_{j\omega_{12}} A_{j\omega_{12}}^\dag\\
+\sum_{\omega,\omega'\in\{\omega_1,\omega_2\}}
[S(\omega,\omega')A_{j\omega'}^\dag A_{j\omega}
+S(-\omega,-\omega')A_{j\omega'} A_{j\omega}^\dag]
\\
+\sum_{\omega,\omega'\in\{0,\pm\omega_0\}}
S(\omega,\omega')A_{j\omega'}^\dag A_{j\omega}
\end{multline*}
[recall that $S(\omega)\equiv S(\omega,\omega)$].

For the baths $j=0,1,2$, we choose the same Drude--Lorentz spectral density $\mathcal J_j(\omega)$ (\ref{EqDrude}). 
The correlation function of the $j$th reservoir can be expressed as
\begin{multline}\label{EqCJ}
C_j(s)=\langle B_j(s)B_j\rangle
\\
=
\int_0^\infty\mathcal J_j(\omega)
\left[
\coth\left(\frac{\beta\omega}2\right)\cos\omega s-i\sin\omega s
\right]d\omega.
\end{multline}
Here $B_j(s)=e^{iH_Bs}B_je^{-iH_Bs}$ and $\langle\cdot\rangle$ denotes the expectation with respect to the thermal state with the inverse temperature $\beta$.

We adopt the high-temperature approximation $\beta\Omega\ll1$. For example, for the used value $\Omega\approx53.08~\rm{cm}^{-1}$ and the minimal considered temperature $T=300$~K, we have $\beta\Omega\approx0.24$. We have used that $\beta=1/k_{\rm B}T$, where $k_{\rm B}\approx0.734~\rm{cm^{-1}/K}$ is the Boltzmann constant. In this case, $\coth(\beta\omega/2)$ in Eq.~(\ref{EqCJ}) can be approximated as $2/(\beta\omega)$ and
\begin{equation}\label{EqC}
C_j(s)\approx\eta\Omega
\left(\frac{2}{\beta\Omega}-i\right)e^{-\Omega s},
\end{equation}
\begin{equation}\label{EqG}
\Gamma_j(\omega)=
\int_0^\infty C_j(s)e^{i\omega s}\,ds
\approx
\frac{\eta\Omega}{\Omega-i\omega}
\left(\frac{2}{\beta\Omega}-i\right).
\end{equation}
So, we have obtained the functions $\Gamma_j(\omega)$ which define both the dissipation rates $\gamma_j(\omega)=2\Re\Gamma_j(\omega)$ and the Lamb shifts of the energy levels $S_j(\omega,\omega')=[\Gamma_j(\omega)-\Gamma^*_j(\omega')]/2i$.

\section{Importance of the refined Lamb-shift Hamiltonian}\label{SecLS}

In this section, we will see that the use of the refined Lamb-shift Hamiltonian (\ref{EqHLS}) instead of the simplified one (\ref{EqHLSsec}) is important in some cases. This is surprising since the Lamb shift is often believed to be insignificant for the dissipation processes at all. Namely, we will consider the case when there is a nontrivial manifold of states stationary for the dissipator, but only one of them commutes with the Hamiltonian. So, the Hamiltonian part and, in particular, the refined Lamb-shift Hamiltonian are crucial for the dynamics toward the unique stationary state in a manifold of states that are stationary for the dissipator only. The existence of two different time scales (dynamics toward a quasi-stationary manifold and dynamics toward the unique final equilibrium) for an open quantum system with nearly degenerate energy levels is rigorously established in Refs.~\cite{MerkliSong,MerkliSongBerman}.

Let us consider the example from the main text and further simplify it: Let two identical qubits interact with each other and with two local dephasing baths. Namely, consider two identical interacting qubits with the Hamiltonian
\begin{equation*}%\label{EqHS}
H_S=E(\sigma_z^{(1)}+\sigma_z^{(2)})+J\sigma_x^{(1)}\sigma_x^{(2)},
\end{equation*}
coupled to two thermal baths of harmonic oscillators. So, the full Hamiltonian is
$$
H=H_S+H_{B,1}+H_{B,2}+H_{I,1}+H_{I,2},$$
where
\begin{equation*}%\label{EqHB}
H_{B,j}=\int_{\mathbb R^d}\omega(k)a_j^\dag(k)a_j(k)\,dk,
\end{equation*}
$a_j(k)$ [$a^\dag_j(k)$] is an annihilation [creation] operator for the $k$th mode of the $j$th bath and $\omega(k)$ is a non-negative real-valued function of $k$. The interaction Hamiltonians are $H_{I,j}=\sigma_z^{(j)}\otimes B_j$, $j=1,2$, where 
$$B_j=\int\,dk\,[\overline g_j(k)a_j(k)+g_j(k) a_j^\dag(k)]$$ 
and $g_j(k)$ are complex-valued functions. That is, in the notations of the main text, we put $\kappa_{jx}=0$, $j=0,1,2$, $\kappa_{0z}=0$, $\kappa_{1z}=\kappa_{2z}=1$, and $E_1=E_2=E$. 

Then, the reduced system dynamics in the subspace spanned by $\{\ket{01},\ket{10}\}$ is independent from that in the subspace spanned by $\{\ket{00},\ket{11}\}$. We choose the initial state as $\rho(0)=\ket{01}\bra{01}$. So, the reduced  system dynamics is confined inside the subspace spanned by $\{\ket{01},\ket{10}\}$ and, thus, is reduced to the two-dimensional case. Let us restrict our consideration to this subspace. Then, the system Hamiltonian is
\begin{equation}\label{EqHSJ}
\widetilde H_S=J(\ket{01}\bra{10}+\ket{10}\bra{01}).
\end{equation}
Its eigenvalues are $\pm J$ with the corresponding eigenvectors $\ket{\pm}=(\ket{01}\pm\ket{10})/\sqrt2$. The interaction Hamiltonians are then $\widetilde H_{I,j}=(-1)^jZ\otimes B_j$, where $$Z=\ket{01}\bra{01}-\ket{10}\bra{10}=
\ket{+}\bra{-}+\ket{-}\bra{+}.$$ 
We again choose the Drude--Lorentz spectral density (\ref{EqDrude}) for both baths  with the coupling strength $\eta=1~\rm{cm}^{-1}$ and the cut-off frequency $\Omega^{-1}=100~\rm{fs}$ ($\Omega\approx53.08~\rm{cm}^{-1}$). The temperatures of the baths are $T_1=T_2=300~\rm{K}$. The parameter $J$ of the system Hamiltonian is $J=2~\rm{cm}^{-1}$.

Now we should choose a proper decomposition of the system Hamiltonian into the reference and perturbation parts. Since $J$ is small with respect to the bath relaxation rate $\Omega$ and comparable to the dissipation constant $\eta$, we treat the whole system Hamiltonian (\ref{EqHSJ}) as small (i.e., the free dynamics takes place on the same time scale as the dissipation): $\widetilde H_S\equiv\lambda^2\widetilde H_S$. In other words, $\widetilde H_S^{(0)}=0$ and $\widetilde{\delta H}_S=\widetilde H_S$.
As we mentioned in the main text, this limiting regime is equivalent to the singular coupling limit.

The unified master equation has the form
\begin{equation}\label{EqDimerUni}
\dot\rho(t)=-i\lambda^2[\widetilde H_S+H_{\rm LS},\rho(t)]+
\lambda^2\mathcal D[\rho(t)],
\end{equation}
where 
\begin{equation}
H_{\rm LS}=S(2J)\ket+\bra+
+
S(-2J)\ket-\bra-
\end{equation}
and
\begin{equation}\label{EqDimerD}
\mathcal D\rho=
\gamma(0)
(Z\rho Z-\rho),
\end{equation}
(we have used that $Z^2=1$). Here $\gamma(\omega)=2\Re[\Gamma_1(\omega)+\Gamma_2(\omega)]$ and $S(\omega)=\Im[\Gamma_1(\omega)+\Gamma_2(\omega)]$ .

The ``simplified'' version of the unified master equation (due to a simpler form of the Lamb-shift Hamiltonian) is
\begin{equation}\label{EqDimerUnisimp}
\dot\rho(t)=-i\lambda^2[\widetilde H_S+\tilde H_{\rm LS},\rho(t)]+
\lambda^2\mathcal D[\rho(t)],
\end{equation}
where 
\begin{equation}\label{EqDimerHLSsec}
\tilde H_{\rm LS}=S(0)(\ket+\bra+
+\ket-\bra-).
\end{equation}

\begin{figure}[t]
\begin{centering}
%\vskip 4mm
\includegraphics[width=\columnwidth]{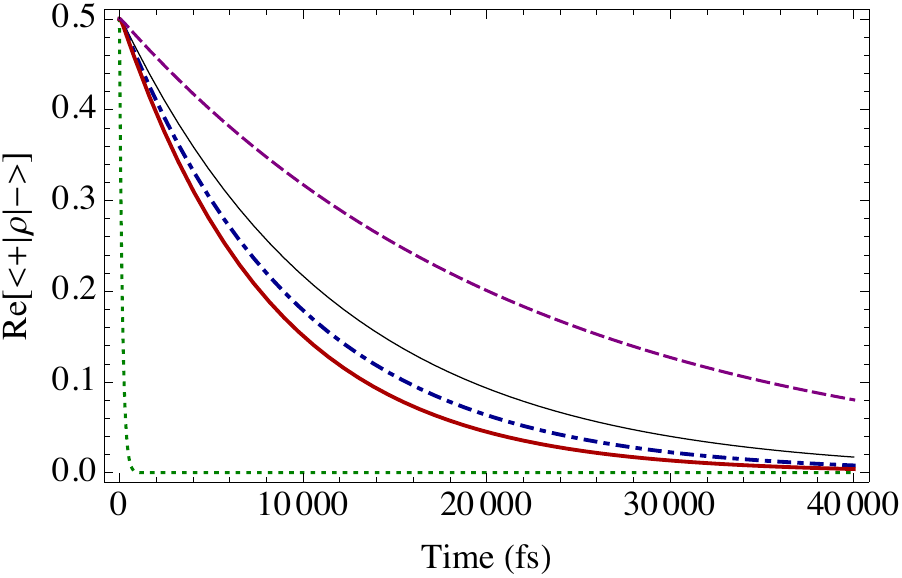}
\vskip -4mm
\caption
{\small
Comparison of calculations of the dynamics of the two-qubit system   according to the numerically exact HEOM method (black solid line), the Redfield equation (blue dash-dotted line), secular (Davies) master equation (green dotted line), the unified master equation (red solid line), and the simplified unified master equation (with the simplified Lamb-shift Hamiltonian), which, in this case, coincides with the known singular coupling master equation (purple dashed line). The plot shows the dynamics of an off-diagonal element (in the eigenbasis of the system Hamiltonian). The secular master equation largely overestimates the rate of decoherence, while the simplified unified master equation significantly underestimates it. The Redfield equation and the unified master equation have good agreement with the numerically exact result.
}
\label{Fig3}
\end{centering}
\end{figure}

A comparison of performances of master equations (\ref{EqDimerUni}) and (\ref{EqDimerUnisimp}) with the numerically exact result of the hierarchical equations of motion (HEOM) (also with the high-temperature approximation \cite{IFl}) is shown on Fig.~\ref{Fig3}. We see that the secular master equation  highly overestimates the rate of decoherence, while the simplified unified master equation significantly significantly underestimates it. The Redfield equation and the unified master equation have good agreement with the numerically exact result.

\begin{remark}
Note that, in this case, the simplified version of the unified master equation (\ref{EqDimerUnisimp}) coincides with the master equation derived in the singular coupling limit \cite{AL,BP,RH,Palmer,Spohn1980,AccFriLu}. So, we have obtained an improved version (\ref{EqDimerUni}) of this master equation, which, as we see, is crucial in some cases.
\end{remark}

%!!!
Let us make some analytic explanations of these numerical results. From the expression of the dissipator (\ref{EqDimerD}), we can see that a peculiarity of this system is that the initial state $\rho(0)=\ket{01}\bra{01}$ as well as any state diagonal in the ``local'' basis $\{\ket{01},\ket{10}\}$ is stationary for the dissipator. However, only the state $I/2$, where $I$ is the identity operator [i.e., diagonal in both the local basis and the eigenbasis (global basis)], is stationary for both the dissipator and the Hamiltonian part. Due to this, the Hamiltonian part and, in particular, the Lamb-shift Hamiltonian are crucial for the dynamics toward the unique stationary state in the manifold of states stationary for the dissipator.

But, as we see from Eq.~(\ref{EqDimerHLSsec}), the simplified approach (the standard singular coupling limit) eliminates the Lamb-shift Hamiltonian: Since only differences between the energy levels matter, the addition of the same quantity $S(0)$ to the levels has no impact.

The initial populations of the eigenstates of $H_S$ are equal: $\braket{\pm|\rho(0)|\pm}=1/2$. The equal populations are stationary since they correspond to the same energy level of the reference system Hamiltonian (equal to zero). So, the dynamics manifests itself only in the decoherence in the eigenbasis of $H_S$. Denote $x(t)=\braket{+|\rho(t)|-}$ and $y(t)=\braket{-|\rho(t)|+}$. Then, the unified  approach gives such equations for the coherences:
\begin{subequations}\label{EqRhopmmp}
\begin{eqnarray}
\dot x&=&-i[2J+S(2J)-S(-2J)]x+\gamma(0)(y-x),\qquad
\label{EqRhopm}\\
\dot y&=&+i[2J+S(2J)-S(-2J)]y-\gamma(0)(y-x).
\label{EqRhomp}
\end{eqnarray}
\end{subequations}
The secular approach treats the energy levels as well separated and thus neglects the coherence--coherence transfer between $\braket{+|\rho(t)|-}$ and $\braket{-|\rho(t)|+}$. Namely, it neglects the terms $+\gamma(0)y$ and $+\gamma(0)x$ in the right-hand sides of Eqs.~(\ref{EqRhopm}) and~(\ref{EqRhomp}), respectively. Doing so, the secular approximation overestimates the decoherence rate. 

The simplified unified approach corresponds to the replacement of $S(\pm 2J)$ by $S(0)$. So, the terms with $S$ vanish. If the signs of $J$ and $S(2J)-S(-2J)$ coincide, such neglect decreases the difference of the rotation rates for the coherences (in the complex plane) and thus decreases the decoherence rate because, as we noticed above, the initial state is stationary for the dissipator and the dynamics is caused by the unitary rotation (followed by the dissipation).

The same fact can be expressed in the language of eigenvalues of the system~(\ref{EqRhopmmp}). Indeed, the smallest eigenvalue in magnitude is 
$$-\gamma(0)+\sqrt{\gamma(0)^2-[2J+S(2J)-S(-2J)]^2}.$$
If $\gamma(0)$ is greater than the second term under the square root, then it can be approximated by
$$-{[2J+S(2J)-S(-2J)]^2}/{2\gamma(0)}.$$
Anyway, the neglection of the terms $S(\pm 2J)$ decreases this eigenvalue whenever the signs of $J$ and $S(2J)-S(-2J)$ coincide. Moreover, if $S(2J)-S(-2J)>2J$, such neglect may cause a large error.  

Recall that both $J$ and $S(\omega)$ are small. In the formal theory, both are multiplied by $\lambda^2$, so, $S(2J)-S(-2J)$ is of the order $\lambda^4$. Hence, in the limit $\lambda\to0$, the effect of this term disappears. This is confirmed by the numerical results if we multiply $J$ and $\eta$ by an infinitesimal dimensionless parameter $\lambda^2$. Thus, Eq.~(\ref{EqThSimp}) is true. However, for the concrete chosen parameters, the simplified equation (the standard master equation for the singular coupling limit) gives an incorrect prediction for the decoherence rate, while the equation with the refined Lamb-shift Hamiltonian provides good results.

Summarizing, the refined Lamb-shift Hamiltonian is important for the case when there exists a nontrivial manifold of states stationary for the dissipator, but only one of them is stationary also for the Hamiltonian part (i.e., commutes with the Hamiltonian). In this case, the Hamiltonian part is crucial for relaxation to this unique state. If we are not in such situation, a simplified Lamb-shift Hamiltonian also can be used. Note that the considered example is practically important because this is a model of excitation energy transfer in a molecular dimer \cite{IFl}. Finally, it is worthwhile to note that a bit higher precision of the Redfield equation in comparison with the unified GKLS master equation seen on Fig.~\ref{Fig3} is achieved at the cost of small violation of positivity on the initial short time (not seen on the plot).

\end{document}